\documentclass[10pt]{article}
\usepackage{graphicx}

\begin{document}

\begin{centering}

{\Large {\bf Acceleration of the Universe in Type-$0$ 
Non-Critical Strings }}

\vspace{0.1in}
{\bf N.~E.~Mavromatos } \\ {\it Department of Physics, Theoretical
Physics, King's College London,\\ Strand, London WC2R 2LS, United
Kingdom.} \\

\vspace{0.1in}
 {\bf Abstract}

\end{centering}

{\small {\it I review 
cosmology within the framework of type-0 non-critical 
strings~\cite{dgmpp}. The instabilities of the 
tachyonic backgrounds, due to the absence of space-time supersymmetry, are 
treated in this framework as a necessary ingredient to ensure 
cosmological flow. 
The model involves 
D3 brane worlds,
whose initial quantum fluctuations induce the non criticality.
I argue 
that this model is
compatible with the current astrophysical observations pointing towards
acceleration of the Universe.
A crucial r\^ole for the correct ``phenomenology'' of the model,
in particular  the order-one value of the deceleration parameter,
is played by the relative magnitude of the flux of the 
five form of the type-0 string as compared to 
the size of the volume of five of the extra dimensions, transverse 
to the direction of the flux-field.}}

\paragraph{} 
Recently there is some preliminary experimental 
evidence
from type Ia supernovae data~\cite{evidomegal},  
which supports the fact that our 
Universe {\it accelerates}
at present: distant supernovae (redshifts  $z \sim 1$) 
data indicate a slower rate of expansion, as compared
with that inferred from data pertaining to nearby supernovae.
Distant supernovae look dimer than they should be, if the 
expansion rate of the Universe would be constant.
If the data have been interpreted right this means 
that $70\%$ of the present energy 
density   
of the Universe consists of an unknown substance 
(``{\it dark energy}''), which, 
for the fit of \cite{evidomegal,evidenceflat},
has been taken to be the standard cosmological constant. 
This last statement, in turn, would imply that our Universe would be  
eternally accelerating (de Sitter), according to standard 
cosmology~\cite{carroll}.
Such evidence 
is still far from being confirmed, but it is 
reinforced by combining these data with 
Cosmic Microwave background (CMB) data (first acoustic peak) 
implying 
a 
spatially $\Omega _{\rm total} =1.0 \pm 0.1$ flat 
Universe~\cite{evidenceflat}.

An important phenomenological parameter, which 
is of particular interest to astrophysicists 
is the {\it deceleration parameter q} of the Universe~\cite{carroll}, 
which is defined as: 
\begin{equation}
q = -\frac{(d^2a_E/dt_E^2)~a_E}{({da_E/dt_E})^2} 
\label{decel}
\end{equation} 
where $a_E(t)$ is the Robertson-Walker scale factor of the Universe, 
and the subscript $E$ denotes quantities 
computed in the so-called Einstein frame, 
that is where the gravity action has the canonical Einstein
form as far as the scalar curvature term is concerned. 
This distinction is relevant in string-inspired effective 
theories with four-dimensional Brans-Dicke type scalars,
such as dilatons, which will be dealing with here. 

It should be mentioned that in standard Robertson-Walker cosmologies
with matter the deceleration parameter can be expressed in terms of the 
matter and vacuum (cosmological constant) energy densities, $\Omega_M$ and $\Omega_\Lambda$ respectively, as follows:
$q=\frac{1}{2}\Omega_M - \Omega_\Lambda $
For the best fit 
Universe~\cite{evidomegal}  
one can then infer 
a present-era deceleration 
parameter 
$q_0 = -0.55 < 0$,  
indicating that the Universe accelerates today with an order one
deceleration parameter. 

If the data have been interpreted right, 
then there are three possible explanations~\cite{carroll}:

\par (i) Einstein's General Relativity is incorrect, and hence 
Friedman's cosmological solution as well. This is unlikely,
given the success of General Relativity and of the 
Standard Cosmological Model in explaining a plethora of other
issues. 

\par (ii) the `observed' dark energy and the acceleration of the Universe 
are due to an `honest' cosmological {\it constant} $\Lambda$ in 
Einstein-Friedman-Robertson-Walker cosmological 
model. This is the case of the best fit Universe 
which matches the supernova and CMB data. 
In that case one is facing the 
problem of eternal acceleration, for the following reason: 
let $\rho_M \propto a^{-3}$ the matter density in the Universe,
with $a (t)$ the Robertson-Walker scale factor, and $t$ the 
cosmological observer (co-moving) frame time. The vacuum energy 
density, due to $\Lambda$, is assumed to be constant in time,
$\rho_\Lambda $ =const. Hence in conventional 
Friedmann cosmologies one has: 
\begin{equation}
\Omega_\Lambda /\Omega_M = 
\rho_\Lambda / \rho_M \propto a(t)^{3}~, 
\label{conventional}
\end{equation}
and hence eventually
the vacuum energy density component will dominate over matter.
{}From Friedman's equations then, one observes that the Universe 
will eventually enter a de-Sitter phase, 
in which $a(t) \sim e^{\sqrt{\frac{8\pi G_N}{3}\Lambda}t}$, 
where $G_N$ is the gravitational (Newton's) constant. 
This implies eternal expansion and acceleration, and most importantly
the presence of a {\it cosmological horizon} 
\begin{equation}
  \delta = a(t) \int_{t_0}^\infty \frac{c dt}{a(t)} < \infty 
\label{cosmhorizon} 
\end{equation}
It is this last feature in de Sitter Universes that 
presents problems in defining proper {\it asymptotic states}, and thus 
a consistent scattering matrix
for field theory in such backgrounds~\cite{eternal}. 
The analogy of such global horizons
with microscopic or macroscopic black hole horizons in this respect
is evident, the important physical difference, however, being that 
in the cosmological de Sitter case the observer lives ``inside'' the horizon,
in contrast to the black hole case. 

Such eternal-acceleration Universes are  
bad news for critical string 
theory~\cite{eternal}, 
due to the fact that strings are by definition theories
of on-shell $S$-matrix and hence, as such, can only accommodate backgrounds
consistent with the existence of the latter.

\par (iii) the `observed effects' are due to the existence of a 
{\it quintessence} field $\varphi$, which has not yet relaxed in its
absolute minimum (ground state), given that the relaxation time 
is longer than the age of the Universe. Thus we are still in a 
non-equilibrium situation, relaxing to equilibrium gradually.
In this drastic explanation, the vacuum energy density, due to the 
potential of the field $\varphi$ will be time-dependent. 
In fact the data point towards a $1/t^2$ relaxation, with $t \ge 10^{60}$
in Planck units, where the latter number represents the age of the 
observed Universe.

It is this third possibility that we have attempted to adopt
in a proper non-critical string theory framework~\cite{ddk,aben} 
in ref. \cite{dgmpp}. 
Non-critical strings can be viewed as non-equilibrium 
systems in string theory~\cite{emn}. 
The advantage of this non-equilibrium
situation lies on the possibility of an eventual exit
from the de Sitter phase, which would allow proper 
definition of a field-theory scattering matrix, thus avoiding the 
problem of eternal horizons mentioned previously.
Exit from de Sitter inflationary phases 
cannot be accommodated (at least to date)
within the context of critical strings~\cite{emn,emnsmatrix}. 
This is mainly due to the fact that
such a possibility requires time-dependent backgrounds in string theory,
which are not well understood within the conformal $\sigma$-model setting. 
On the other hand, there is sufficient evidence that such 
a `graceful exit possibility' from the inflationary 
phase can be realized in non-critical
strings, with a time-like signature of the Liouville field, which thus 
plays the r\^ole of a Robertson-Walker comoving-frame 
time~\cite{emn,emnsmatrix}.
The evidence came first from toy two-dimensional specific 
models~\cite{grace}, 
and was extended in \cite{dgmpp} to four-dimensional models 
based on the so-called type-0 non-supersymmetric strings~\cite{type0}.

The latter string theory has four-dimensional brane worlds, whose fluctuations 
have been argued in \cite{dgmpp} to lead to super criticality
of the underlying string theory, necessitating Liouville dressing
with a Liouville mode of time-like signature. In general, Liouville 
strings become critical strings after such a dressing procedure,
but in one target-space dimension higher. 
However in our approach~\cite{emn,dgmpp}, instead of increasing the 
initial number of target space dimensions ($d=10$ for type-0 strings), 
we have 
identified the world-sheet zero mode of the Liouville field with the 
(existing) target time. In this way, the time may be thought of as 
being responsible of re-adjusting itself (in a non-linear way), 
once the fluctuations
in the brane worlds occur, so as to restore the disturbed 
conformal invariance
of the underlying world sheet theory. 

One of the most important results of \cite{dgmpp} is the 
appearance of a time-dependent central charge deficit 
in the underlying conformal theory, acting as a vacuum energy density
in the respective target-space lagrangian. This is crucial for 
a `graceful 
exit' from the inflationary de Sitter phase, and the absence of 
eternal acceleration. In fact the asymptotic (in time) theory 
is that of a flat (Minkowski) target-space $\sigma$-model with a 
linear dilaton~\cite{aben} in the string frame. 
In the Einstein frame, i.e. in a redefined
metric background in which the Einstein curvature term in the 
target-space effective action has the canonical normalization, 
the universe is linearly expanding, which is the limiting
case in which the horizon (\ref{cosmhorizon}) diverges logarithmically;
hence such a theory can admit properly defined asymptotic states 
and $S$-matrix amplitudes. The linear dilaton background has 
been shown~\cite{aben} to be a consistent background for string theory,
despite being time dependent,
in the sense of satisfying factorizability (for certain discrete values
of the asymptotic central charge though), modular invariance and 
unitarity.

Another important aspect of our solution 
is the fact that the extra bulk dimensions are compactified
in such a way that one is significantly larger than the others, thereby
leading to effective five-dimensional brane world scenaria. 
This is a consequence of an appropriately chosen five-form flux background.
This feature is one of the most important
ones of the type-0 stringy cosmologies, which are known to be characterized
by the existence of non-trivial flux form fields coming from the 
Ramond sector of the brane worlds~\cite{type0,dgmpp}.

The reader might object to our use of type-0 backgrounds due to 
the existence of tachyonic backgrounds. 
Although at tree level it has been demonstrated that 
the above-described flux forms can stabilize such backgrounds
by shifting away the tachyonic mass poles, however, recently this
feature has been questioned at string loop level.
Nevertheless, in the context of our cosmological model,
such quantum instabilities are expected probably as a result
of the {\it non-equilibrium} nature of our 
relaxing background, in which the time dependent dilaton field
plays the r\^ole of the quintessence field.
Indeed, as demonstrated in \cite{dgmpp} the 
asymptotic in time value of the tachyon background is zero,
and hence such a field disappears eventually from the spectrum,
which is 
consistent with the asymptotic equilibrium nature of the 
ground state.

The effective ten-dimensional target space action
of the type-$0$ String, upon which we base our low-energy analysis, 
reads 
to ${\cal O}(\alpha ')$ in
the Regge slope $\alpha '$~\cite{type0}:
 \begin{eqnarray}\label{actiontype0} S&=&\int d^{10} x
\sqrt{-G}\Big{ [}e^{-2\Phi}\Big{(}R + 4(\partial_M \Phi)^2
-\frac{1}{4}(\partial_M T)^2 - \frac{1}{4}m^2T^2 \nonumber \\&~~ &
-\frac{1}{12}H_{MNP}^2\Big{)} -  \frac{1}{4}(1 + T +
\frac{T^2}{2})|{\cal F}_{MNP\Sigma T}|^2\Big{]} \end{eqnarray}
where capital Greek letters denote ten-dimensional indices, $\Phi$
is the dilaton, $H_{MNP}$ denotes the field strength of the
antisymmetric tensor field, which we shall ignore in the present
work, and $T$ is a tachyon field of mass $m^2 <0$. In our analysis
we have ignored higher than quadratic order terms 
in the tachyon potential. The
quantity ${\cal F}_{MNP\Sigma T}$ denotes the appropriate
five-form of type-$0$ string theory, with non trivial flux, 
which couples to the tachyon
field in the Ramond-Ramond (RR) sector via the function $f(T)=1 +
T + \frac{1}{2}T^2$.

{}From (\ref{actiontype0}) one sees easily the important r\^ole of
the five-form ${\cal F}$ in stabilizing the ground state. Due to
its special coupling with the quadratic $T^2$ term in the Ramond-Ramond (RR)
sector of
the theory, it yields an effective mass term for the tachyon which
is positive, despite the originally negative $m^2$
contribution~\cite{type0}.
As mentioned previously, such a stability has recently been 
questioned in the context of string loop corrections,
but as we have mentioned previously this is rather a desirable
feature of the approach, in view of the claimed
cosmological instabilities. 

As argued in \cite{dgmpp} 
{\it fluctuations of the brane worlds} involved in 
the construction of type-0 string theory result in 
{\it supercriticality} of the underlying $\sigma$-model,
with inevitable consequence 
the addition of the following 
term to the action
(\ref{actiontype0}) \begin{equation} \label{qterm}
 - \int d^{10} x
\sqrt{-G}e^{-2\Phi} Q(t)^2 
\end{equation}
where $Q(t)$ is the 
central-charge deficit of the non-equilibrium non conformal 
$\sigma$-model theory. The time here is identified with the 
(world-sheet zero mode of) the Liouville field, 
and the $t$ dependence of the central charge deficit is in accordance
with the concept of a Zamolodchikov C-function~\cite{zam}, 
a ``running central charge'' of a 
non-conformal theory, interpolating between two conformal 
(fixed point) theories. 
The sign of $Q(t)^2$  is positive if one assumes
supercriticality of the string~\cite{aben,ddk,grace,dgmpp}, 
which is the case of the model of \cite{dgmpp}. 
It is important to remark that 
in general, $Q^2(t)$ depends on the $\sigma$-model
backgrounds fields, being the analogue of Zamolodchikov's
$C$-function~\cite{zam}. 
As explained in \cite{dgmpp}, the explicit 
time dependence of $Q(t) $ reflects the existence 
of relevant operators in the problem,
other than the background fields considered in (\ref{actiontype0})
which are treated collectively in the present context.
Such operators have been argued to represent {\it initial} 
quantum fluctuations of the brane world. A plausible scenario, for instance, 
would be that the initial disturbance that takes the system out of equilibrium
is due to an impulse on the D3 brane worlds coming from either
a scattering off it of a macroscopic number of closed string bulk states
or another brane in scenaria where the bulk space is uncompactified 
(e.g. ekpyrotic universes {\it etc.}~\cite{ekpyrotic}).
For times long after the event, memory of the details of this process is
kept in the temporal evolution of $Q^2(t)$, which is 
determined self-consistently by means of the Liouville 
equations, as we shall see below.

The ten-dimensional metric configuration we considered in \cite{dgmpp} 
was: 
\begin{equation}
G_{MN}=\left(\begin{array}{ccc}g^{(4)}_{\mu\nu} \qquad 0 \qquad 0 \\
0 \qquad e^{2\sigma_1} \qquad 0 \\ 0 \qquad 0 \qquad
e^{2\sigma_2} I_{5\times 5} \end{array}\right)
\label{metriccomp}
\end{equation}
where lower-case Greek indices are four-dimensional space time
indices, and $I_{5\times 5}$ denotes the $5\times 5$ unit matrix.
We have chosen two different scales for internal space. The field
$\sigma_{1}$ sets the scale of the fifth dimension, while
$\sigma_{2}$ parametrize a flat five dimensional space. In the
context of cosmological models, we are dealing with here, the
fields $g_{\mu\nu}^{(4)}$, $\sigma_{i},~i=1,2$ are assumed to
depend on the time $t$ only.

As we demonstrated in \cite{dgmpp}, a consistent background choice
for the flux form field will be that in which the flux is parallel to 
to the fifth dimension $\sigma_1$. This implies actually 
that the internal space is crystallized (stabilized) 
in such a way that this dimension is much larger than the 
remaining five $\sigma_2$. 

Upon considering the fields to be time dependent only, 
i.e. considering spherically-symmetric homogeneous backgrounds, 
restricting
ourselves to the compactification (\ref{metriccomp}), and assuming
a Robertson-Walker form of the \\ four-dimensional metric, with scale
factor $a(t)$, the generalized conformal invariance conditions
and the Curci-Pafutti $\sigma$-model renormalizability constraint~\cite{curci}
imply a set of differential equations, which we solved numerically 
in \cite{dgmpp}.  The system can also be solved analytically
in the phase of large times after the initial fluctuations,
where the various fields obey linearized approximations.

The generic form of the Liouville equations,
which express restoration of conformal invariance after 
Liouville dressing, reads~\cite{ddk,emn,schmid,dgmpp}:
\begin{equation} 
  {\ddot g}^i + Q(t){\dot g}^i = -{\tilde \beta}^i 
\label{liouvilleeq}
\end{equation} 
where ${\tilde \beta}^i$ are the Weyl anomaly coefficient of the 
stringy $\sigma$-model on the background $\{ g^i \}$. 
In the model of \cite{dgmpp} the set of $\{ g^i \}$ contains
graviton, dilaton, tachyon, flux and moduli fields $\sigma_{1,2}$ 
whose vacuum expectation values control the size of the extra dimensions.
As mentioned above, these equations should be supplemented by
the Curci-Paffuti renormalizability condition~\cite{curci}.

As argued in \cite{dgmpp} such equations 
correspond to solutions of equations of motion 
derived from a
ten-dimensional effective action. This is an important
and non-trivial consequence of the gradient flow property of the 
$\sigma$-model ${\tilde \beta}^i$ functions, according to which:
${\tilde \beta}^i = {\cal G}^{ij}\frac{\delta {\cal F}[g]}{\delta g^j} $
where the flow functional  ${\cal F}[g]$ is essentially 
the target-space effective 
action, depending on the background configuration under consideration. 

An equivalent set of equations (in fact at
most linear combinations) come out from the corresponding
four-dimensional action after dimensional reduction. Of course
this reduction leads to the string 
or $\sigma$-model
frame, in which there are dilaton exponential 
factors in front of the Einstein term in the action. 
We may turn to the Einstein frame,
in which such factors are absent, and the 
Einstein term is canonically normalized.
In this latter frame, 
the cosmological time is defined by
$dt_E =  e^{-\Phi+\frac{ \sigma _1 + 5 \sigma _2}{2}}dt $.
The extra dimensions freeze out quickly~\cite{dgmpp} 
and their constant values can be absorbed in a redefinition
of the Newton's constant~\cite{dgmpp}. 
This yields the Einstein-frame time in the form: 
\begin{equation}\label{einsteintime} 
t_E=\int ^t e^{-\Phi (z)}dz 
\end{equation}  
In this late-time phase, the Einstein-frame dilaton 
varies logarithmically with the Einstein time $t_E$ (\ref{einsteintime}) 
\begin{equation} 
\Phi _E = {\rm const} -{\rm ln}t_E~,
\label{einsteindil} 
\end{equation}
while the Einstein-frame 
``vacuum'' energy is related to the central charge deficit as~\cite{dgmpp}:
\begin{equation} 
\Lambda_E = e^{2\Phi}Q^2(t) 
\label{vacuumenergy} 
\end{equation} 
in Planck $M_P$ units. 
As discussed in \cite{dgmpp}, due to the non-equilibrium nature of the 
non-critical string Universe, which has not yet relaxed to its 
ground state, $\Lambda_E$ 
should be considered rather as an effective potential,
in much the same way as the potential of a (non-equilibrium)
quintessence field, whose r\^ole is played here the 
dilaton $\Phi$~\cite{emnsmatrix,dgmpp}.

\begin{figure}[h]
\centering
\includegraphics[scale=0.5]{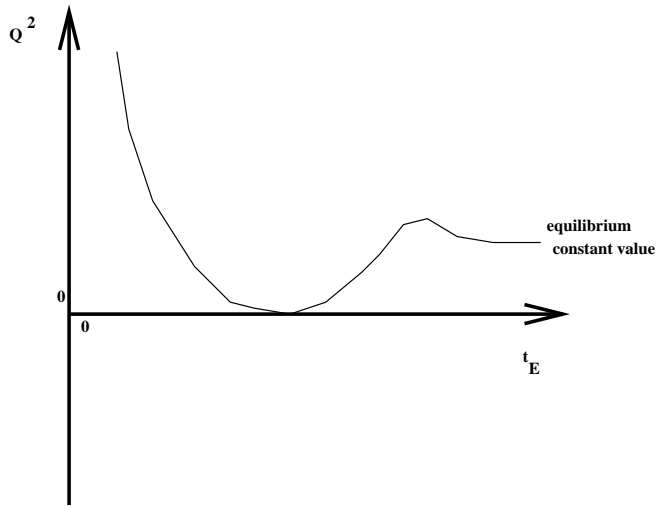}
\caption {The evolution of the central charge deficit $Q^2$
in the Einstein frame. Immediately after inflation, 
the deficit passes through a phase where it 
first vanishes, and then oscillates before 
relaxing to an equilibrium constant value asymptotically.} 
\label{centralcharge:fig}
\end{figure}

\begin{figure}[h]
\centering
\includegraphics[scale=0.5]{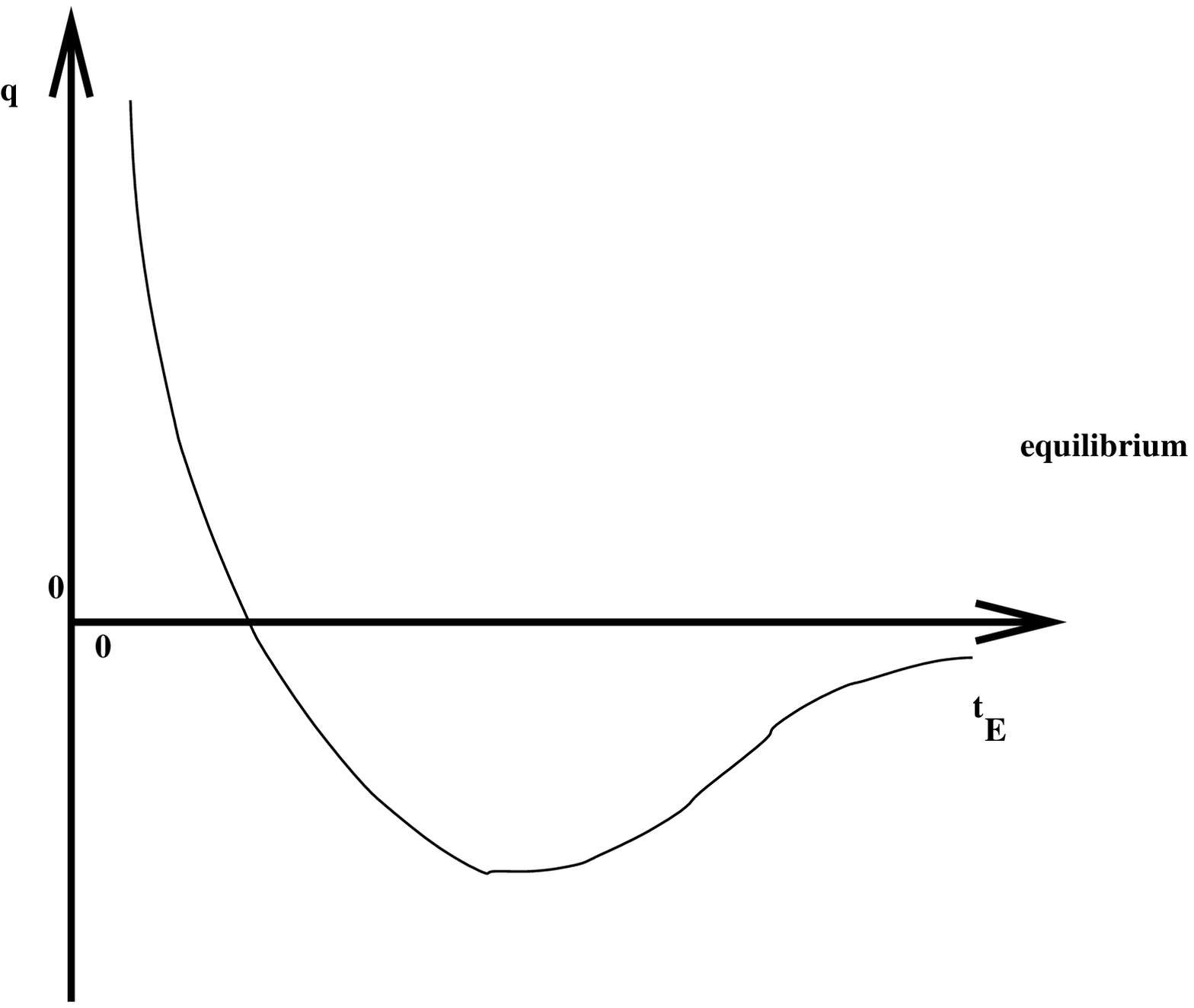}
\caption {The evolution of the decelerating parameter $q$  
of the type-0 string Universe in the Einstein frame.}
\label{decel:fig}
\end{figure}

The results of the analysis of \cite{dgmpp} 
are summarized in figs. \ref{centralcharge:fig} and
\ref{decel:fig}.
The numerical solution 
is supported by analytical considerations
for the asymptotic field modes (late cosmological-frame times $t \to \infty$).
Our solution 
demonstrates  
that the scale factor of the Universe, 
after the initial singularity, enters a
short inflationary phase and then, in a smooth way, goes into 
a flat
Minkowski spacetime with a linear dilaton for 
asymptotically long times $t \rightarrow \infty$
in the $\sigma$-model frame.
Note that this is a consistent $\sigma$-model background~\cite{aben}  
in the sense of satisfying modular invariance and factorization
of $S$-matrix elements. This is actually one of the most important 
points of our work in \cite{dgmpp}, namely that 
there is a smooth exit from the de Sitter phase in such a way
that one can {\it appropriately define} asymptotic on-shell 
states, and hence an $S$-matrix.

We observe from fig. \ref{decel:fig} that, 
after its graceful exit from the inflationary phase, 
the type-$0$ non-critical string Universe 
passes first through a decelerating phase, 
which is then succeeded by an accelerating
one, 
before the Universe relaxes asymptotically to is steady state (equilibrium)
value.

From fig. \ref{centralcharge:fig} we also observe 
that, 
immediately after inflation, the central charge deficit $Q^2$ 
passes through a
{\it metastable point} where it vanishes. 
This has to do with the fact 
that at this point 
the square root $Q$ changes sign.
After this point the central charge deficit 
oscillates (as a function of time) before it relaxes to its constant asymptotic
value, $q_0$, 
which should be taken as one of the values for which the conformal theory of 
\cite{aben} is valid. The oscillatory nature is consistent with the 
time-like signature of the Liouville mode as explained in \cite{dgmpp}.

\begin{figure}[h]
\centering
\includegraphics[scale=0.5]{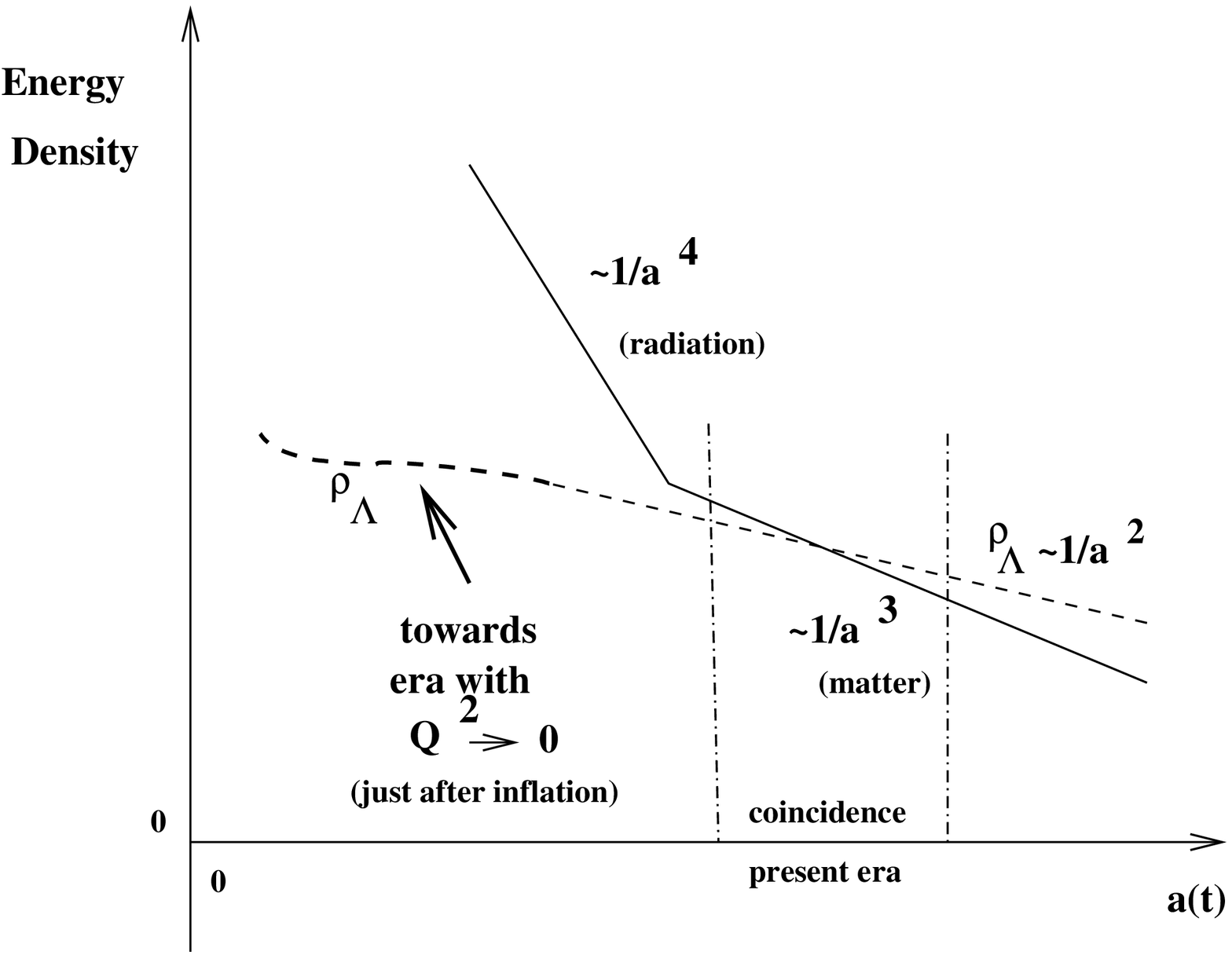}
\caption {The evolution of the energy densities of matter, radiation
and of the quintessence field (dilaton) vs. the scale factor 
of the Universe  
in the Einstein frame. At early stages the energy density of the 
quintessence field decreases significantly, as compared with the 
rest, and the coincidence situation is lost. This is due to the 
behaviour of the central charge deficit of the model, shown 
in figure \ref{centralcharge:fig},
which dives in to zero for a short period 
immediately after inflation.} 
\label{energydensity:fig}
\end{figure}

This behaviour of the central charge 
deficit is important in determining the 
evolution of the energy densities of the type-$0$ 
string Universe, which are depicted in fig. \ref{energydensity:fig}. 
We should remark that 
in the type-$0$ string/brane scenario, ordinary matter may be assumed
attached to the brane, and hence purely four dimensional. 
The latter is going to resist the deceleration 
of the universe, according to standard arguments, but it is not expected
to change the order of magnitude of such quantities as 
the Hubble parameter and the vacuum energy, determined in \cite{dgmpp} in the 
absence of matter.
In this sense one may obtain the observed `coincidence situation' of the 
present era, where the matter and `dark energy' contributions 
are roughly of the same order of magnitude~\cite{carroll}. 
In our scenario it is the time dependence of both `dark energy' and
`matter contribution', in conjunction 
with the value the time $t_E$ has at present, 
roughly $t_E \sim 10^{60}M_P^{-1}$,
that is held responsible for this coincidence
situation.

In our model 
the tracking (coincidence) of the matter energy density by that of the 
quintessence dilaton field 
is a feature only of the present era, which is a welcome feature 
phenomenologically.  
As the time $t_E$ elapses further, 
the matter contribution will become subdominant,
as scaling like $a_E^{-3}$. For very large
times $t_E$ in the far future, as we have seen above, 
the dominant contributions will be the ones 
due to the 
non-constant in time 
`dark energy component' 
$\Lambda (t_E) \sim a_E^{-2}$, which asymptotes to zero,
as the system reaches its equilibrium value.  
This makes a quantitative difference 
in scaling as compared with the standard Robertson-Walker scenario
with a constant vacuum energy.

For large times, such as the present era, 
a straightforward computation yields~\cite{dgmpp}
the Hubble parameter: 
\begin{equation}\label{hubble2} 
H(t_E) \simeq \frac{\gamma^2 t_E}{1 + \gamma^2 t_E^2}
\end{equation}
where the constant $\gamma \propto V_5/|C_5|$, with 
$C_5$ the flux of the five form, which is taken along the fifth 
extra dimension, and 
$V_5$ the volume
of the five extra dimensions, 
transverse to the flux direction. 

The deceleration parameter (\ref{decel}), in the same regime 
of $t_E$, is:
\begin{equation} 
q(t_E) \simeq -\frac{1}{\gamma^2 t_E^2}
\label{decel4}
\end{equation}
Finally, the ``vacuum energy'' can be obtained from (\ref{vacuumenergy}):
\begin{equation} 
\Lambda (t_E) \simeq \frac{q_0^2 \gamma^2}{F_1^2 ( 1 + \gamma^2 t_E^2)}
\label{cosmoconst}
\end{equation}
where $q_0$ is the constant asymptotic value of the central charge deficit 
$Q^2$. In the model of \cite{dgmpp} conformal invariance   
requires that $q_0/F_1 \sim {\cal O}(1)$. 

If, therefore, one {\rm defines} the  
{\it present era} by the time regime
\begin{equation}
\gamma \sim t_E^{-1} 
\label{condition}
\end{equation} 
in the Einstein frame, 
then from (\ref{decel4}) 
it becomes clear that an order one negative 
value of $q$ is obtained, in agreement with the
prelimianry astrophysical observations. 
The important point is that 
this is compatible with large enough times $t_E$ (in string units) 
for 
$|C_5|/V_5 \gg 1~$. 
This condition can be guaranteed
{\it either} for small radii of the five of the extra dimensions 
{\it or} for a large value of the flux $|C_5|$ of the five-form 
of the type-$0$ string. Notice that the relatively large extra dimension,
in the direction of the flux, 
decouples from this condition, thus 
allowing for the possibility of effective five-dimensional models 
with large uncompactified fifth dimension.

\end{document}